\documentclass[fleqn,usenatbib]{mnras}

\usepackage{CJK} 
\usepackage{bm} 
\usepackage{newtxtext,newtxmath}

\usepackage[T1]{fontenc}

\DeclareRobustCommand{\VAN}[3]{#2}
\let\VANthebibliography\thebibliography
\def\thebibliography{\DeclareRobustCommand{\VAN}[3]{##3}\VANthebibliography}

\usepackage{graphicx}	
\usepackage{amsmath}	

\title[Cosmological simulation of FDM subhaloes]{Cosmological zoom-in simulation of fuzzy dark matter down to $z=0$: \\tidal evolution of subhaloes in a Milky Way-sized halo}

\author[H. Y. Jowett Chan et al.]{
Hei Yin Jowett Chan,$^{1}$
Hsi-Yu Schive(\begin{CJK*}{UTF8}{bkai}薛熙于\end{CJK*}),$^{1,2,3,4}$\thanks{E-mail: hyschive@phys.ntu.edu.tw}
Victor H. Robles,$^{5}$
Alexander Kunkel,${^{2}}$
Guan-Ming Su,${^{2}}$ \newauthor
and Pin-Yu Liao$^{3}$
\\
$^1$Physics Division, National Center for Theoretical Sciences, National Taiwan University, Taipei 106319, Taiwan\\
$^2$Department of Physics, National Taiwan University, Taipei 10617, Taiwan\\
$^3$Institute of Astrophysics, National Taiwan University, Taipei 10617, Taiwan\\
$^4$Center for Theoretical Physics, National Taiwan University, Taipei 10617, Taiwan\\
$^5$Department of Physics, Applied Physics \& Astronomy, Rensselaer Polytechnic Institute, Troy, NY 12180, USA}

\date{Accepted XXX. Received YYY; in original form ZZZ}

\pubyear{\the\year{}}

\begin{document}
\label{firstpage}
\pagerange{\pageref{firstpage}--\pageref{lastpage}}
\maketitle

\begin{abstract}

Subhaloes are critical in distinguishing dark matter models, yet their evolution within galactic haloes, particularly in the Fuzzy Dark Matter (FDM) model, remains challenging to fully investigate in numerical simulations. In this work, we employ the fluid-wave hybrid scheme recently implemented in the \texttt{GAMER-2} code to perform a cosmological zoom-in simulation of a Milky Way-sized halo with an FDM particle mass of $m=2\times10^{-23}~\mathrm{eV}$. It simultaneously resolves the solitonic core of the host halo and tracks the complex tidal evolution of subhaloes down to redshift $z=0$. We examine the internal structure of subhaloes by analyzing their density profiles, velocity dispersions, and density power spectra across various redshifts. Our findings show that partially tidally stripped subhaloes deviate from the core-halo mass relation; their solitons remain intact and are enveloped by smaller granules predominantly from the host halo. Furthermore, our simulation unravels a complex tidal evolution of FDM subhaloes. On the one hand, we observe a subhalo core undergoing complete tidal disruption at $z \sim 0.14$, which later reemerges near the outskirts of the host halo around $z \sim 0$. This disruption event, characterized by a core contaminated with interference fringes from the host halo's wave function, occurs earlier than previously predicted. On the other hand, FDM subhaloes have denser cores before infall due to the presence of central solitons, making them more resilient to tidal disruption than their N-body counterparts. Our results demonstrate \texttt{GAMER-2}'s capability to resolve non-linear FDM substructure down to $z=0$, paving the way for future studies of larger FDM subhalo samples with heavier particle masses. The simulation code \texttt{GAMER-2} and the simulation setup used in this work are available at \url{https://github.com/gamer-project/gamer}.
\end{abstract}




\section{Introduction} \label{sec:intro}
Numerical simulations of dark matter stand as the main way to predict the process of structure formation within the non-linear regime. For instance, cosmological simulations of the Lambda Cold Dark Matter ($\Lambda \mathrm{CDM}$) model reveal that dark matter produces substructures in self-gravitating haloes, including subhaloes and stellar streams. These substructures allow us to test dark matter models against observed properties of dwarf galaxies in the local Universe. The well-known small-scale challenges to the $\Lambda\mathrm{CDM}$ model, including the missing satellites, core-cusp and too-big-to-fail problems, all arise from the comparison studies between simulations of CDM subhaloes and dwarf galaxies (see, for instance, \citet{Bullock:2017xww} for a review). 

In the past two decades, simulation groups started including galaxy formation physics in cosmological simulations, demonstrating that baryonic processes can suppress the formation of dwarf galaxies, and thus provide potential solutions to these small-scale challenges \citep{Sawala2016,Wetzel2016,Kimmel2019}. However, new tensions are also emerging \citep{Boruk2022,Owan2015}, including research that suggests an opposite conclusion, a too many satellite problem \citep{Kelly2019,honma2024}. It remains uncertain whether the $\Lambda$CDM model can offer a fully consistent explanation for the observed properties of dwarf galaxies within the Milky Way. Therefore, dwarf galaxies, which are closely linked to dark matter substructures, remain to serve as a powerful dark matter probe. 

\begin{figure*}
    \centering
    \includegraphics[scale=1.2]{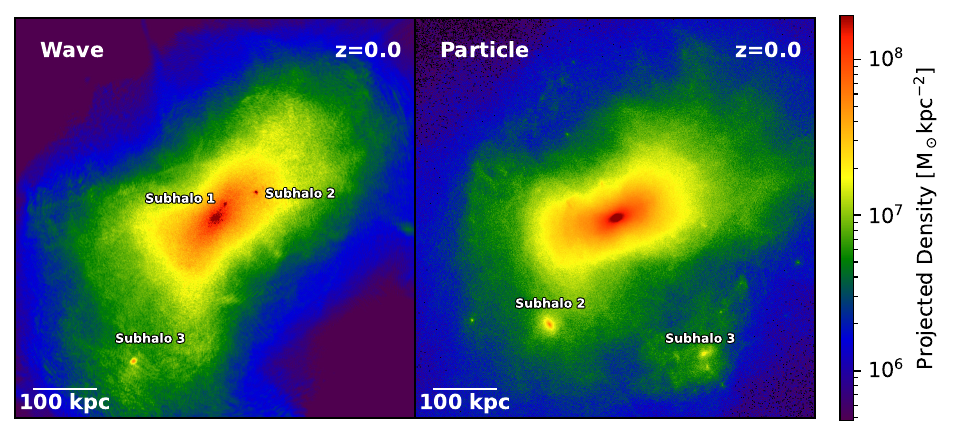}
    \caption{Comparison between the projected density of a Milky Way-sized halo of the FDM model with a particle mass $m=2\times 10^{-23}~\mathrm{eV}$ on the left and 
    N-body simulation with an FDM initial condition on the right at $z=0$. Since both simulations start with the same initial condition of the FDM model, the differences in the density projections arise from the distinction between wave-like and particle-like non-linear dynamics. The subhaloes formed at similar positions in both simulations before infall. By tracking their evolution, we can identify the counterparts of each subhalo after infall in the two simulations. 
    }
    \label{fig:MWz0}
\end{figure*}

In fact, similar comparison tests have been extensively applied to alternative dark matter models, such as the Warm Dark Matter (WDM) and Self-Interacting Dark Matter (SIDM) models. Due to their heavier particle mass, these models, like the CDM model, belong to the set of so-called particle-like dark matter models \citep{Hui:2016ltb}. Consequently, much effort has been invested in developing more sophisticated numerical techniques for the N-body simulations to predict accurate statistics and inner structure of subhaloes at high resolution within a large simulation volume (see, for instance, \citet{Angulo2022} for a review). 

In contrast, substructures of the FDM model remain less understood compared to particle-like dark matter models. The FDM model, as a subset of the Ultralight Dark Matter model, proposes extremely light bosonic dark matter particles, with a mass of $m \sim 10^{-22}$--$10^{-20}~\mathrm{eV}$. The small particle mass leads to wave-like behavior on astrophysical scales, such as the formation of a solitonic core within each dark matter halo and the suppression of small-scale structures below the de Broglie wavelength \citep{Hu2000,Schive2014a}. On large scales, the FDM and CDM models agree well, explaining why the FDM model is considered a candidate for resolving the small-scale challenges of the CDM model. Previous studies have constrained the FDM particle mass by comparing statistics and the inner structure of FDM subhaloes with dwarf galaxies \citep{Marsh2019,Gonzalez-Morales:2016yaf,Chen:2016unw,Safarzadeh2019,Calabrese:2016hmp,Kohei2021,Nadler2021,Dalal2022}. However, a precise understanding of the FDM subhaloes requires self-consistent large-scale cosmological FDM simulations, which are yet to be achieved. Therefore, those constraints were derived relying either on semi-analytical models \citep[e.g.][]{Du:2016aik} or simulations that could not fully capture the non-linear dynamics simultaneously, including dynamical friction, granule heating, tidal stripping, and interaction with a Milky Way-sized host halo.

The numerical difficulty of performing FDM simulations is reflected by the fact that a Milky Way-sized halo at $z=0$ is still difficult to achieve, even for particle mass smaller than $10^{-22}~\mathrm{eV}$. For example, the large-volume FDM simulations by \citet{Simon2021}, performed in a comoving $(14.85~\mathrm{cMpc})^3$  box using a pseudospectral method on a uniform grid with $N = 8192^3$ points, could not resolve de Broglie wavelength beyond $z=3$. Consequently, different groups have employed different algorithms aiming to overcome the challenge of resolving small-scale wave dynamics while preserving the large-scale structure in coarser resolution \citep{Mina2020,Bodo2022,Nori2021}.  

Zoom-in simulations represent a method to invest most of the computational resources into a region of interest, by increasing the resolution only in a sub-volume of the entire simulation box. This technique has been broadly used for particle-like dark matter models, such as CDM, WDM, and SIDM, since they follow a similar equation of motion. However, it is still rarely applied to the Ultralight Dark Matter model.

In this work, we demonstrate that using the recently developed novel hybrid scheme in the \texttt{GAMER-2} code \citep{GAMER2, Kunkel2024}, we can successfully perform zoom-in simulations of a Milky Way-sized FDM halo with substructures for a particle mass of $m=2\times 10^{-23}~\mathrm{eV}$ (see Fig. \ref{fig:MWz0}) down to $z=0$. Our simulation simultaneously resolves the solitonic core of the host halo and three subhaloes in-falling from the surrounding filaments, demonstrating the strength of the hybrid scheme
, while revealing the complex tidal evolution of FDM subhaloes within the host. Although the simulation of this work assumes a particle mass lower than the canonical values $m \sim 10^{-22}$--$10^{-20}~\mathrm{eV}$, the simulation serves as an important stepping stone toward FDM zoom-in simulations with larger particle mass. The paper is organized as follows: In Section \ref{section:method}, we introduce the governing equation of motion of the FDM model in a hybrid formulation as well as the simulation set-up for a cosmological FDM zoom-in simulation.  In Section \ref{section:result_host}, we introduce the host halo alongside a convergence analysis. In Section \ref{section:result_subhalo}, we analyze the properties of the accreted FDM subhaloes. The conclusion in Section \ref{section:conclusion} discusses avenues for future work based on our results.

\section{Simulation set-up}
\label{section:method}
\subsection{Hybrid Formulation}
Our simulation is performed using the \texttt{GAMER-2} code \citep{GAMER2} on a GPU cluster using $8$ computing nodes, each equipped with an AMD Ryzen Threadripper PRO 5975WX 32-Cores CPU and an NVIDIA GeForce RTX 3080 Ti GPU. The total computational time to reach $z=0$ is approximately $8$ days. The details of the recent upgrade of the hybrid scheme are presented in \citet{Kunkel2024}, and here we briefly summarize the numerical technique for evolving the wave function of the FDM model. The governing equation of motion of the FDM model is the Schr\"odinger-Poisson equation in comoving coordinates,
\begin{align}
    ia^2\partial_t \psi(\bm{x}, t) &= \left(-\frac{\hbar}{2m}\nabla^2 + \frac{m}{\hbar} \phi(\bm{x}, t)\right) \psi(\bm{x}, t), \label{eq:ComovingSchroedinger}\\
    \nabla^2 \phi(\bm{x}, t) &= 4\pi Ga(|\psi(\bm{x}, t)|^2 - \rho_\mathrm{b}) \label{eq:ComovingPoisson},
\end{align}
where $\hbar$ is the reduced Planck constant, $m$ is the FDM particle mass, $a$ is the scale factor, $G$ is the gravitational constant, and $\rho_\mathrm{b}$ is the comoving background density. 

On large scales where the density field is smooth and velocities are high, it is advantageous to treat the density field $\rho$ and the phase field $S$ in $\psi = \sqrt{\rho} \exp(iS)$ as fundamental variables and discretise the kinetic operator in the fluid formulation of the Schr\"odinger equation: the Hamilton-Jacobi-Madelung equations in comoving coordinates
\begin{align}
    \frac{a^2m}{\hbar} \partial_t   \rho +  \nabla \cdot \left(\rho \nabla S\right)  &= 0, \label{eq:BohmContinuity}\\
    \frac{a^2m}{\hbar} \partial_t S +  \frac{1}{2}\left(\nabla S\right)^2 + \frac{m^2}{\hbar^2} \phi - \frac{1}{2} \frac{\nabla^2 \sqrt{\rho}}{\sqrt{\rho}} &= 0. \label{eq:BohmPhase}
\end{align} 
The fluid formulation is valid as long as the density does not vanish. In regions of strong destructive interference, where density voids form and the fluid formulation fails, we switch to the wave formulation of the Schr\"odinger equation Eq. (\ref{eq:ComovingSchroedinger}) and evolve the kinetic operator using a $13^\mathrm{th}$-order accurate local pseudospectral method relying on Gram-Fourier extensions \citep{Lyon2009}, which can achieve higher spatial accuracy than the previously adopted $6^\mathrm{th}$-order finite difference method \citep{Schive2014a}. This is important as the finite difference method requires a much higher spatial resolution than the hybrid scheme to accurately represent the high infall velocity near massive haloes. Compared to existing hybrid methods for the FDM model \citep{velmatt2020, Bodo2022}, our numerical approach allows a direct conversion between fluid and wave representations without approximation of the phase field, while also accounting for the quantum pressure term. Additionally, our local pseudospectral method in the refined wave region is significantly more accurate than conventional finite difference methods.

In practice, the entire simulation volume is mostly evolved by the fluid formulation at high redshifts. Once interference patterns begin to emerge in the overdense structures, a part of the simulation volume will switch to the wave scheme in refined regions, while the remaining part of the simulation domain with a weak quantum pressure continues to use the fluid scheme (see Fig. \ref{fig:fwm}). The hybrid scheme allows us to conduct simulations with large box sizes, which is essential to study a Milky Way-sized halo because the simulation volume must be large enough to enclose the necessary mass for forming a $10^{12}~M_\odot$ halo at $z=0$. In addition, a large simulation volume avoids numerical artifacts due to the periodic boundary conditions.   

\subsection{Zoom-in Simulation}
\begin{figure}
    \centering
     \hspace*{-1.2cm}     \includegraphics[scale=0.35] {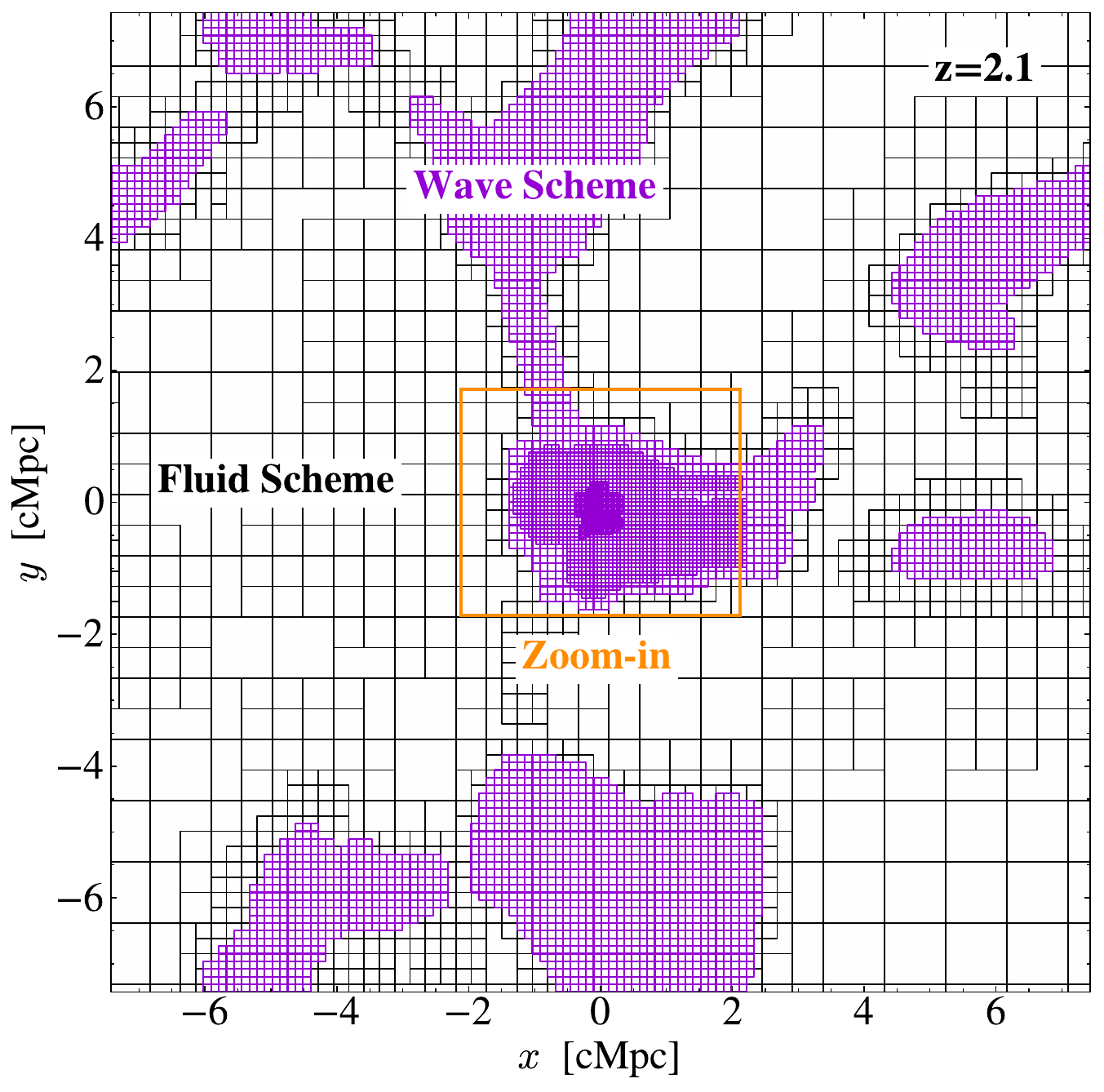}
    \caption{The figure shows the AMR grid distribution of a zoom-in FDM simulation using the hybrid scheme, where the zoom-in box has dimensions of $4.2~\mathrm{cMpc} \times3.4~\mathrm{cMpc}\times3.4~\mathrm{cMpc}$ at $z=2.1$. The fluid scheme, which solves the Hamilton-Jacobi-Madelung equations (Eqs. (\ref{eq:BohmContinuity}) and (\ref{eq:BohmPhase})), is active in refinement levels 0 to 2 represented by black grids. The wave scheme, which solves the Schr\"odinger equation (Eq. (\ref{eq:ComovingSchroedinger})), is active in refinement levels 3 to 7 represented by purple grids. The orange box highlights the zoom-in box that surrounds the target Milky Way-sized halo at the center and allows refinement up to level 7. In contrast, the region outside of the zoom-in box only allows refinement up to level 3. The highest refinement level for the fluid and wave scheme has a resolution of $14.66~\mathrm{ckpc}$ and $0.46~\mathrm{ckpc}$, respectively.}
    \label{fig:fwm}
\end{figure}
The hybrid scheme is implemented into the \texttt{GAMER-2} code, allowing us to take advantage of the built-in adaptive mesh refinement (AMR) algorithm (see Fig. \ref{fig:fwm}). We can further speed up the simulation by limiting refinement to a region of interest, a technique called zoom-in that is explained in the following. 

The initial condition of our simulation begins with a uniform resolution of $256^3$ grid points at base-level 0 in a $(14.85~\mathrm{cMpc})^3$ volume\footnote{Throughout the paper, we present comoving length units as ckpc and cMpc, while physical length units are denoted as kpc and Mpc.}. As the simulation evolves, regions will be refined if they satisfy any of the following three refinement criteria: the density criterion, the Madelung refinement criterion, and the spectral refinement criterion \citep{Kunkel2024}. The Madelung refinement criterion is designed to detect regions with destructive interference while ensuring sufficient spatial resolution required for the density and phase fields in the fluid formulation. The spectral refinement ensures that the mesh resolution for the wave formulation is properly resolving the de Broglie wavelength. 

The maximum density allowed by the density criterion in the fluid region is $64\rho_\mathrm{b}$, where $\rho_\mathrm{b}$ is the comoving background density. However, in practice, the density in the fluid region may not reach this limit because destructive interference can form at densities lower than $64\rho_\mathrm{b}$, thereby triggering the Madelung refinement criterion. As described above, density is not the only criterion determining the transition between the fluid and wave formulation. Instead, the Madelung criterion serves as a more reliable criterion to control this transition, as it triggers refinement before strong destructive interference forms. Virialized haloes form after shell-crossing, and we observe wave interference emerging at the shell-crossing regions. This indicates that the Madelung criterion has already converted these regions from the fluid to the wave formulation at the first shell-crossing, prior to halo virialization. Consequently, all subhaloes in our simulation form in the wave region.
\begin{figure*}
    \centering
    \includegraphics[scale=0.9]    {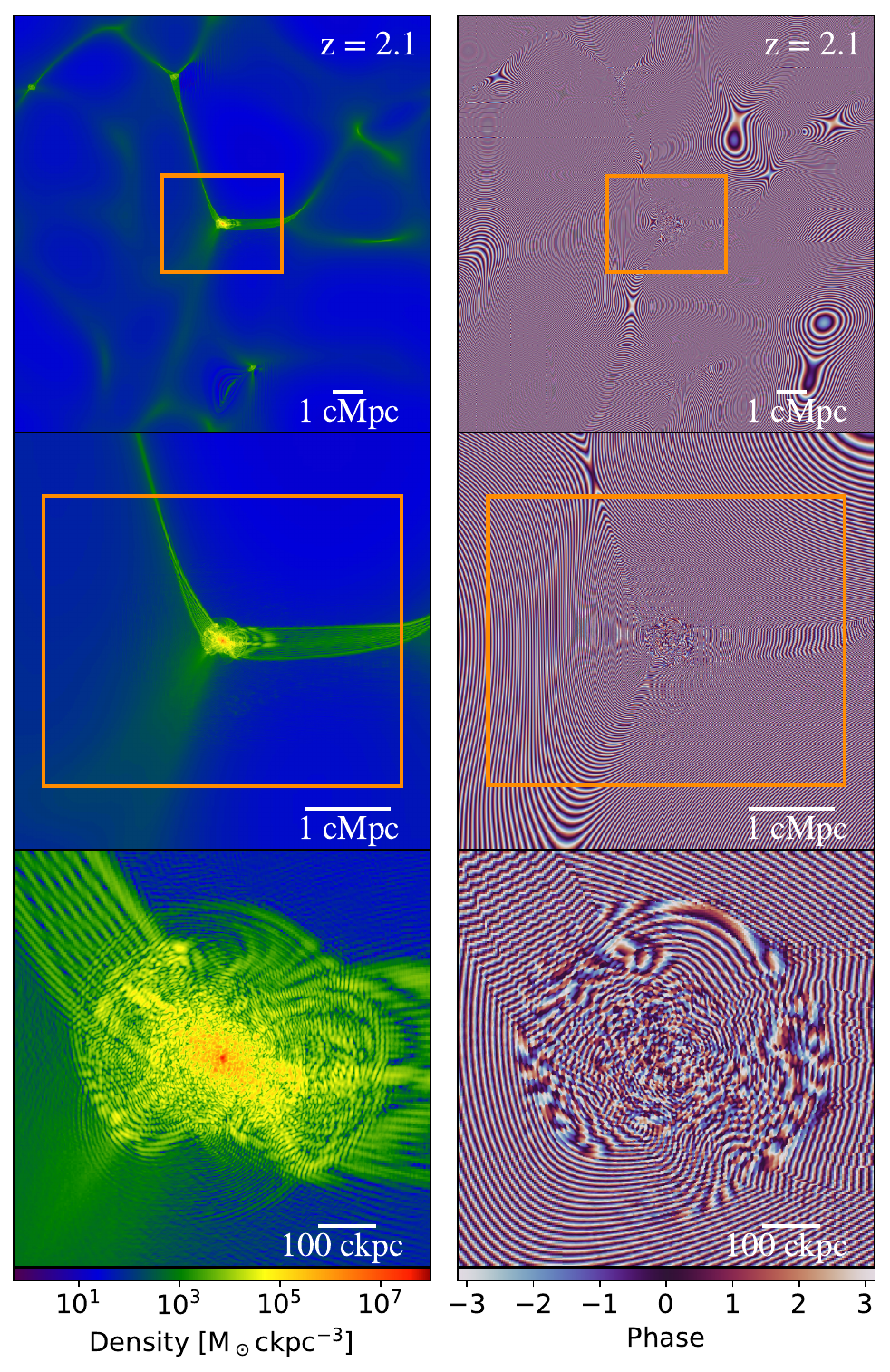}
    \caption{A sequential zoom-in of the density distribution (left column) and phase distribution (right column) into the target halo at $z=2.1$. The orange line highlights the border of the zoom-in region. Both density and phase are smooth, even at the fluid-wave interface, thanks to the hybrid scheme that solves the boundary matching problem to handle the transition of phases at the fluid-wave interface. The corresponding mesh distribution is shown in Fig. \ref{fig:fwm}.}
    \label{fig:fwm_zoom}
\end{figure*}

In particular, our simulation allows a maximum of $7$ 
refinement levels, going from the base-level $0$ to level $7$, where each level doubles the spatial resolution. This leads to an effective resolution of $32768^3$ grid points or $0.46~\mathrm{ckpc}$ on level $7$\footnote{To avoid confusion, we remind that the resolution of the refinement level is not directly equal to the power of base $2$. For instance, level 7 in this work refers to an effective spatial resolution of $(2^{8+7})^3=32768^3$ grid points rather than $(2^{7})^3$, where $2^8$ is the base-level resolution.}. The zoom-in technique comes into play when we impose a rectangular zoom-in box that further restricts refinement away from the halo of interest. For regions outside of the zoom-in box, the maximum refinement level is restricted to level 3, whereas the zoom-in region is refined up to the maximum level 7. The zoom-in region is defined using the Lagrangian volume of a pre-selected $2\times 10^{12}~\mathrm{M}_\odot$ halo at $z=0$ obtained from an N-body simulation (see right panel of Fig. \ref{fig:MWz0}) using \texttt{Gadget-2} \citep{Gadget2} with the same initial condition. Since we can trace back halo member particles in the N-body simulation to earlier redshifts, we select the outermost member particles to define the time-dependent Lagrangian volume as the zoom-in region for the FDM simulation.

We remind that an N-body zoom-in simulation only requires defining a Lagrangian volume at the starting redshift. It is because the N-body simulation itself is based on a Lagrangian numerical approach, so higher resolution particles will dynamically redistribute themselves to track the halo of interest at lower redshifts. In comparison, the hybrid scheme is an Eulerian approach that solves the equation of motions on fixed meshes. So we need to manually specify the Lagrangian volumes at different redshifts in order to mimic the time-dependent evolution of the Lagrangian regions in the N-body simulation.

In addition, we further extend each side of the zoom-in box by $ 0.74~\mathrm{cMpc}$, which is motivated by two reasons: First, the wave dynamics of the FDM model introduces quantum pressure that counteracts gravity, delaying the formation of FDM haloes. For example, we observe the core structure having a lower density at a higher redshift than its N-body counterpart that follows the NFW profile, as shown in the bottom left panel of Fig. \ref{fig:MWsliceprof}. As a result, we expect the infall speed of self-gravitating FDM systems to be lower than particle systems, especially at higher redshifts. Second, the extension can prevent contamination of the target halo by spurious overdensities created at the boundary of the zoom-in region due to the sharp gradient of the spatial resolution. For instance, at lower redshifts, the zoom-in box is mostly surrounded by grids at refinement level 3, whereas the region within the zoom-in box is filled with refinements at levels 6 or 7. Without the extension, the insufficient spatial resolution at level 3 outside the zoom-in box could contaminate the target halo.  

Fig. \ref{fig:fwm_zoom} demonstrates the density and phase fields of the hybrid simulation, sequentially zooming into the region of interest. We see that both density and phase are smooth even at the fluid-wave interface. This is because the hybrid scheme solves the forward and backward boundary matching problem at the interface, allowing precise reconstruction of the wave function from the fluid variables and vice versa (see \citet{Kunkel2024}). Again, the difficulty of FDM simulations is due to the highly oscillating wave function, as clearly shown in the phase field of Fig. \ref{fig:fwm_zoom}. Note that high-velocity structures, corresponding to short wavelengths, exist not only in the local high-density region but also in the more extensive low-density region. Therefore, the combined AMR algorithm and hybrid scheme are critical for performing FDM simulations with larger box sizes to capture non-linear structures on smaller scales. 

In summary, we set up the hybrid zoom-in simulation by applying the fluid formulation from refinement levels $0$ to $2$, the wave formulation from levels $3$ to $7$, and restricting the maximum level outside the zoom-in box to level 3 (see Fig. \ref{fig:fwm}). In particular, the last configuration allows applying the wave formulation outside the zoom-in box, which is essential while combining the hybrid scheme with the zoom-in technique, as explained in the following.

\subsubsection{The fluid-wave interface at zoom-in boundaries}

At high redshifts, most of the simulation volume is evolved using the fluid representation. But at lower redshifts, parts of the entire simulation domain and most of the zoom-in region are evolved using the wave formulation. If the boundary of the zoom-in region is an interface between the fluid and wave formulations, the fluid scheme can fail due to the infamous incapability of the Madelung equation of handling destructive interference and vortices. For example, the high velocities in regions of destructive interference within filaments will require prohibitively small time steps. We stress that the fluid-wave interface is generally not an issue for simulations without a zoom-in region because regions exhibiting destructive interference and vortices will always switch to a wave-wave interface. However, this is not necessarily true for zoom-in simulations. For example, if the maximum refinement level outside the zoom-in region is restricted to level 2, the boundary of the zoom-in box will be surrounded by a fluid-wave interface at lower redshifts. This occurs because the region outside cannot switch to the first wave level, which is at level 3. Consequently, the fluid scheme will be forced to use prohibitively small time steps to handle the high-velocity destructive interference structure at its neighboring wave grids, which stalls the simulation.

To avoid this problem, we reserve at least one wave level for regions outside of the zoom-in box. In this way, the hybrid scheme can handle the destructive interference at the zoom-in boundary by switching to the wave solver outside of the zoom-in box, as demonstrated in Fig. \ref{fig:fwm}. It is evident that such an issue arises uniquely in FDM hybrid zoom-in simulations but not in an N-body zoom-in simulation.

\subsection{Initial condition}
We construct the initial condition of the cosmological zoom-in FDM simulation at $z=100$ on a $256^3$ uniform mesh with a comoving side length of $14.85~\mathrm{cMpc}$ by providing \texttt{MUSIC} \citep{MUSIC} with the truncated linear power spectrum generated by \texttt{axionCAMB} \citep{AxionCAMB} using a particle mass $m=2\times 10^{-23}~\mathrm{eV}$. The cosmological parameters are $h=0.67$, $\Omega_m=0.316$, $\Omega_\Lambda=0.684$, and $\sigma_8=0.8119$ \citep{Planck2018}. 
As mentioned in the previous section, unlike an N-body zoom-in simulation, it is not required to provide the Lagrangian patch to MUSIC since the zoom-in technique is fully handled by the \texttt{GAMER-2} code. Moreover, the lack of small-scale power in the initial power spectrum of a small particle mass suggests that a uniform mesh is sufficient to resolve the initial density and velocity fields of the FDM model. The conversion from the velocity field to the phase field is performed by solving the Poisson problem $\nabla^2S=(m/\hbar)(\nabla\cdot\mathbf{v})a^2$, where $\mathbf{v}=\partial \mathbf{x}/\partial t$ and $\mathbf{x}$ is the comoving spatial coordinate. Note that the physical velocity is given by $\mathbf{v}_\mathrm{phy}=(\hbar/m)\boldsymbol{\nabla} S /a + Ha\mathbf{x}$, where $H$ is the Hubble parameter.

\section{Milky Way-sized Host Halo}
\label{section:result_host}

\begin{figure*}
    \centering
    \includegraphics[scale=0.8]{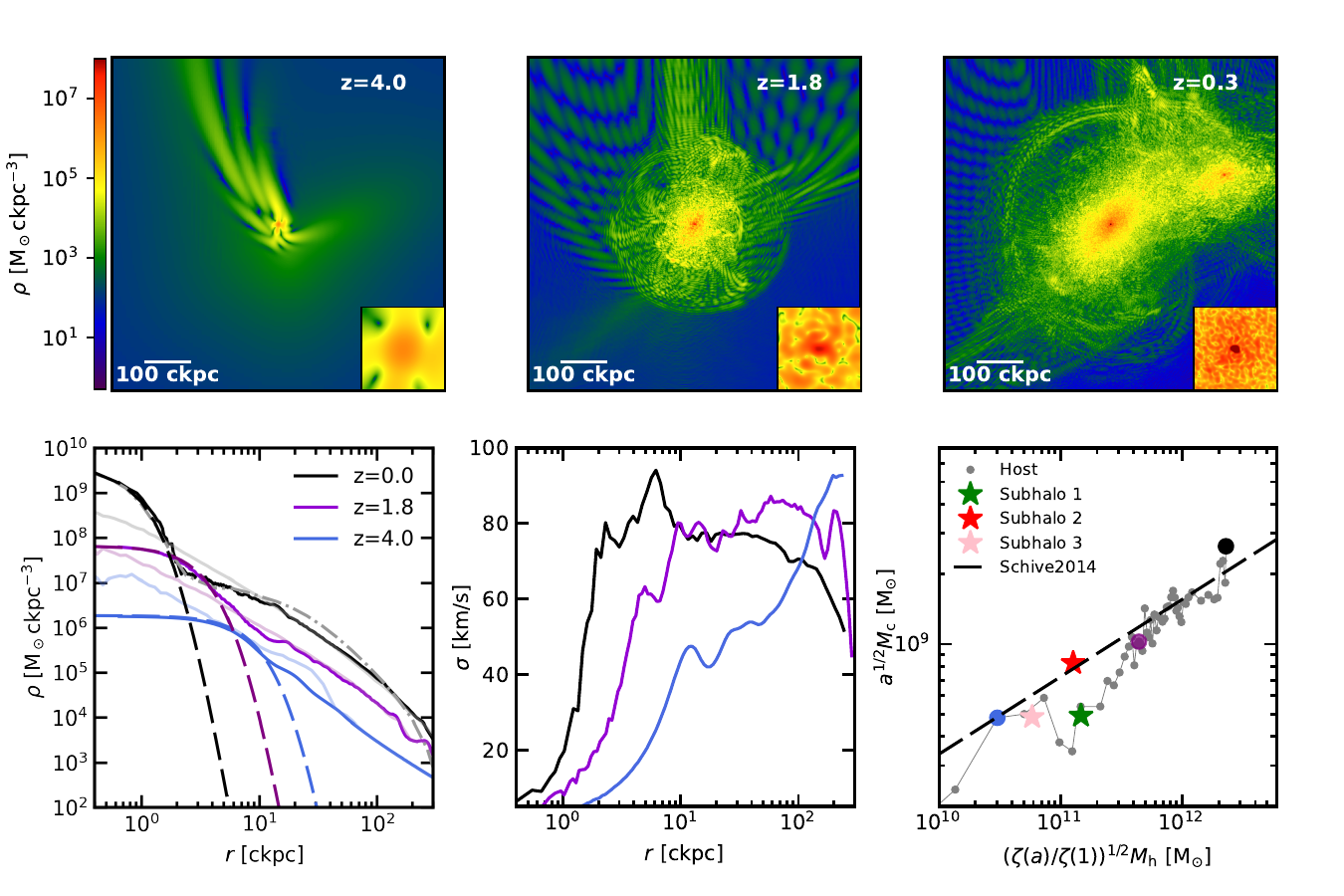}
    \caption{The top row shows the time evolution of density slices through a Milky Way-sized FDM halo at $z=4.0, 1.8$, and $0.3$. The subplots zoom further into the soliton of the host with a $37~\mathrm{ckpc}$ side box. We can see concentric mass shells at $z=1.8$ located at $r\sim100$--$150~\mathrm{ckpc}$ and a subhalo being accreted at $z=0.3$. The solitonic core becomes manifest after $z\lesssim2.$ The bottom row shows the density profiles (bottom-left), velocity dispersion profile (bottom-middle), and core mass-halo mass relation (bottom-right) of the Milky Way-sized halo at various redshifts. Note that profiles at $z=0.3$ and $0.0$ are similar so we highlight the $z=0.0$ results in the bottom row. In the bottom-left panel, solid lines are profiles from the FDM simulation. Dashed lines are the analytical soliton profiles from Eq. \ref{eq:core-density}. Light solid lines are profiles from the N-body simulation with the same initial condition. The grey dot-dashed line is an analytically reconstructed FDM halo based on the eigenstate approach. In the bottom-right panel, grey circles are the core-halo mass relation of the Milky Way-sized host halo at different redshifts, while the colored circles highlight specific redshifts corresponding to the left and middle panels. When the soliton of the host halo becomes more stable after $z\lesssim2$, it closely follows the core-halo relation of \citet{Schive2014b} (black dashed line). The colored stars are the core-halo mass relation of the subhaloes before being accreted into the host, which all lie close to the relation of \citet{Schive2014b}. See text in Section \ref{section:result_host} for details.}
    \label{fig:MWsliceprof}
\end{figure*}

Fig. \ref{fig:MWsliceprof} shows the density slices of the time evolution of the host halo with a virial mass of $2\times10^{12}~\mathrm{M}_\odot$ at $z=0$. We define the halo virial mass as $M_\mathrm{h} = (4\pi r_\mathrm{h}^3/3) \zeta(z) \rho_\mathrm{b}$ \citep{Bryan1998}, where $r_\mathrm{h}$ is the halo virial radius, $\rho_\mathrm{b}$ is the background matter density and $\zeta \sim 180$ ($350$) for $z = 0$ ($\ge 1$). 
In the upper panels of Fig. \ref{fig:MWsliceprof}, it is clear that the wave interference and the core structure, as the unique features of the FDM halo, are both well resolved in our zoom-in simulation for the entire Milky Way-sized halo.  More importantly, the non-linear structure and evolution of the halo is fully captured in the simulation, including the growth of the host halo by accreting mass from the filaments, the concentric mass shells formed by the first apocenters of infalling matter shown at the $z=1.8$ panel at $r \sim 100$--$150~\mathrm{ckpc}$, and the accretion of subhaloes shown at the $z=0.3$ panel. All of the mentioned structure formation were not captured by any previous wave-based simulations since their target haloes are either not massive enough or unable to reach $z=0$, resulting in the lack of substructures. The host halo in our simulation accreted three subhaloes in total, 
two of which survived until $z=0$, and one of which was tidally disrupted but re-emerged at $z=0$, as shown in the left panel of Fig. \ref{fig:MWz0}. We will discuss it further in Section \ref{section:result_subhalo}.

To verify if the observed non-linear structures are physical or numerical artifacts, we simulate an N-body counterpart using \texttt{Gadget-2} \citep{Gadget2} with the same initial condition. The resulting density projection of the simulation is shown in the right panel of Fig. \ref{fig:MWz0}. Some of those substructures exist in the N-body counterpart. However, due to the lack of quantum pressure in the non-linear regime, there is no direct mapping between the subhaloes of FDM and N-body simulations, which is clearly seen by the difference in the position of the subhaloes between the two simulations. Moreover, subhalo 1 is tidally disrupted in the N-body simulation but survives in the FDM simulation, confirming that N-body simulations with FDM initial conditions cannot fully reproduce the structure formation of FDM subhaloes. We further measure and compare the density profiles of the host halo in both simulations, shown in the bottom left panel of Fig. \ref{fig:MWsliceprof}. At all redshifts, the density profiles of the FDM halo show a cored structure, distinctly different from that of the N-body counterpart. However, more crucially for confirming numerical convergence, the outer regions of the FDM profiles align well with the profiles obtained from the N-body simulation. Such an agreement between the FDM and N-body simulations, even at high redshifts, has become a standard practice for assessing the numerical convergence of our zoom-in FDM simulations \citep{Liao2024}.  
\begin{figure*}
    \centering
    \hspace*{-1.5cm}
    \includegraphics[scale=1.5]{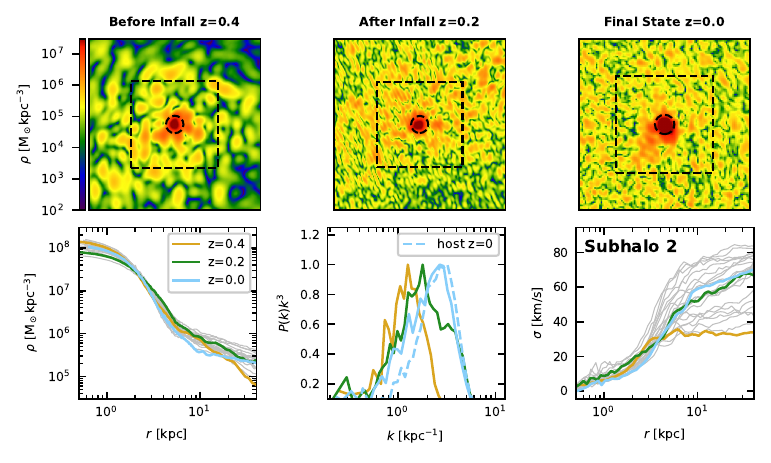}
    \caption{Properties of subhalo 2, which survives with an intact core. The top row shows density slices at redshifts $z=0.4, 0.2$, and $0$, with core radii of $1.34~\mathrm{kpc},1.53~\mathrm{kpc}$, and $1.40~\mathrm{kpc}$, respectively. Data within the box-shaped dashed line with a $20r_\mathrm{c}$ side length and outside of the inner dashed circle with a $2r_\mathrm{c}$ radius are used for the power spectra measurement. The lower-left, lower-middle, and lower-right plots show the density profiles, power spectra, and velocity dispersion of the subhalo at different redshifts, respectively. The grey lines show the remaining profiles at redshifts between $z=0.4$ and $0$ that are not labeled. The dashed line in the lower-middle panel denotes the power spectrum of the host halo. }

    \label{fig:subhalo}
\end{figure*}

In addition, we compare the simulated FDM density profile at $z=0$ to an analytically reconstructed FDM halo based on an eigenstate approach \citep{Lin2018}, denoted as the dot-dashed line in Fig. \ref{fig:MWsliceprof}. While the overall shapes of the profiles are similar, there are small but noticeable deviations between the reconstructed and simulated FDM haloes, possibly due to the assumptions of spherical symmetry or the specific form of the wave distribution function in the eigenstate approach.

Fig. \ref{fig:MWsliceprof} shows that the core structure at the center of the FDM halo can be well described by the analytical soliton profile \citep{Schive2014b},
\begin{equation}
    \label{eq:core-density}
    \begin{aligned}
\rho_{\mathrm{c}}(r) &= 1.9\times10^{9}\, a^{-1} \left( \frac{10^{-23}~\mathrm{eV}}{m} \right)^{2} \\
&\times \left( \frac{\mathrm{ckpc}}{r_{\mathrm{c}}} \right)^4\,\left[1+0.091\left(\frac{r}{r_\mathrm{c}}\right)^2\right]^{-8}\frac{\mathrm{M}_\odot}{\mathrm{ckpc^3}},
    \end{aligned}
\end{equation}
where $r_c$ is the comoving core radius at which density drops to half of the peak value. This solitonic core structure stems from the quantum pressure counteracting the gravitational collapse. Such a quantum effect not only creates a signature in the density field but also in the velocity field. Since the energy content of the soliton is dominated by gravity and quantum pressure energy, we expect the kinetic energy of bulk motion to be minimal within the soliton. This is confirmed in the bottom middle panel of Fig. \ref{fig:MWsliceprof}, showing the velocity dispersion profile at three different redshifts computed by $\sigma = \sqrt{\sum_i \sigma^2_i/3}$ where $i=x,y,z$ and $\sigma_i^2=\langle\rho v^2_i\rangle/\langle\rho\rangle - \langle\rho v_i\rangle^2/\langle\rho\rangle^2$ \citep{Chowdhury2021} for physical $v_i$. All of the velocity dispersion profiles drastically decrease within the core radius.

Importantly, when comparing our velocity profiles with those produced by FDM simulations using the Smoothed Particle Hydrodynamics (SPH) method \citep{Nori2023}, the velocity profiles in SPH simulations show a constant velocity, or a plateau, within the core. This contrasts with the substantially decreasing velocity observed in our simulated Milky Way-sized halo.  Moreover, the density profiles of FDM haloes in \citet{Nori2023} do not show a sharp transition at the outskirt of the core, but such a sharp transition is evident in our Milky Way-sized halo at $z=0$ around $r \sim 2~\mathrm{ckpc}$. The distinction in the velocity dispersion and density profile of the core confirms that the underlying mechanism of core formation differs between the wave-based approach, through numerical solving the Schr\"odinger equation, and the SPH approach, which demands further investigation.

Lastly, the time evolution of the relation between the core mass and halo mass is shown in the right panel of Fig. \ref{fig:MWsliceprof}. Our Milky Way-sized halo closely follows the core-halo relation predicted by \citet{Schive2014b} once the core becomes stable after $z\sim2$. We conclude that the Milky Way-sized halo produces properties that are consistent with previous studies. 

\section{Subhalo}
\label{section:result_subhalo}
Since our FDM simulations are grid-based, the commonly used particle-based subhalo finders are not compatible with our simulations or require substantial changes in the subhalo finder algorithm \citep[e.g.][]{Simon2021}. We only have three subhaloes in the simulation, so we identify the center of each subhalo through a simple density-based criterion.

Both the FDM and N-body simulations contain substructures in the host halo, but the non-linear wave dynamics involve quantum pressure, so the formation history of subhaloes is different. In fact, three subhaloes survive in the FDM simulation, whereas only two survive in the N-body counterpart because one of them is tidally disrupted (see Fig. \ref{fig:MWz0}).

We select subhalo 2 and subhalo 3 (see left panel of Fig. \ref{fig:MWz0}) as our representative FDM subhaloes to showcase the non-linear dynamics between the host and the substructures in a self-consistent FDM simulation. Fig. \ref{fig:subhalo} highlights three particular evolutionary stages of subhalo 2. At $z=0.4$, the subhalo has a virial mass of $1\times 10^{11}~\mathrm{M}_\odot$ and a physical virial radius of $110~\mathrm{kpc}$. It is located outside of the physical virial radius of the host, which is $221~\mathrm{kpc}$ at $z=0.4$, and therefore has not yet been accreted by the host. We define this stage of evolution as "before infall" \footnote{The infall time often refers to the moment when the subhalo crosses the virial radius of the host halo.  For instance, see \citet{Rocha2012}.}. At $z=0.2$, the subhalo is well within the virial radius "after infall" but has not experienced the first pericenter passage. After two pericenter passages, the subhalo reaches its final state at $z=0$. The density slices of subhalo 2 at $z=0.4,0.2$, and $0$ are presented in the top panels of Fig. \ref{fig:subhalo}. Here we first focus on the two main features of subhaloes: the core structure (Section. \ref{sec:core}) and granules (Section. \ref{sec:granules}). A discussion on the abundance of subhaloes is deferred to Appendix. \ref{sec:abundance}.

\subsection{Core structure}
\label{sec:core}
The lower-left panel of Fig. \ref{fig:subhalo} shows that the core densities of subhalo 2 at different redshifts only differ by a maximum of 2. The small difference in the core density implies the core mass is unaffected by the interaction with its host, as confirmed in the top panel of Fig. \ref{fig:subhalo_cdsgrr} showing the evolution of the core mass from $z=0.4$ to $0.0$. Since subhalo 2 has experienced two pericenter passages (see the second row of Fig. \ref{fig:subhalo_cdsgrr}), it must have experienced tidal stripping from the host. As a result, most of the mass loss occurs outside the core of subhalo 2, but its core mass remains largely unchanged. Subhalo 1 exhibits a similar evolutionary history. The intact cores and the stripped halo mass of subhalo 1 and 2 are in alignment with previous idealized simulations \citep{Schive2020} that included an external gravitational potential to mimic the tidal effects of a Milky Way host at an orbital radius of $100~\mathrm{kpc}$.

The results have important implications for the core mass-halo mass relation \citep{Schive2014b,Bodo2016,Mocz2017,Jowett2022,Mina2020,Nori2021} of the FDM model, which suggests that we can infer a core mass for a given halo mass and redshift. 

Although we are not able to provide a quantitative measurement of the core mass-halo mass relation for subhaloes due to the difficulty of measuring the halo mass of the subhaloes during tidal stripping, we can measure the core-halo mass relation of the subhaloes before infall and explore the implication after infall. For instance, subhalo 2 has a halo mass of $\sim 10^{11}~\mathrm{M}_\odot$ and a core mass of $8\times 10^{8}~\mathrm{M}_\odot$ before infall at $z=0.41$, which agrees with the relation of \citet{Schive2014b} (see the bottom-right panel of Fig. \ref{fig:MWsliceprof}). In fact, all three subhaloes agree with the core-halo mass relation of \citet{Schive2014b} before infall. After infall, we expect the core-halo mass relation of subhalo 2 to deviate more and more from the relation of \citet{Schive2014b} since the core mass is unchanged while the halo mass decreases.

The same reasoning can also be applied to subhalo 1 and 3. Therefore, our subhaloes provide strong evidence that the universal core mass-halo mass relation does not apply to subhaloes that have experienced tidal stripping because the tidally stripped subhalo remains to have the same core mass after infall. At most, the core mass-halo mass relation of \citet{Schive2014b} serves as the upper limit of subhalo mass for a given core mass. Empirical relations commonly used in the community are all derived from isolated haloes, which explains why such non-universality is difficult to be fully revealed by previous studies. For particle mass heavier than the value adopted in our simulation, $m=2\times10^{-23}~\mathrm{eV}$, solitons become denser and more resistant to tidal stripping and disruption. Therefore, we expect the core mass to remain constant after infall even for heavier particle mass, suggesting that the discrepancy with the universal core mass-halo mass relation likely persists.

\subsubsection{Core Disruption}
Tidal disruption of FDM subhaloes was expected to be enhanced by a "quantum tunneling" effect \citep{Hui:2016ltb}, where subhalo mass can tunnel through the potential barrier at the tidal radius. \citet{Du2018} provided a more detailed demonstration of the tunneling effect through ideal simulations. When the outer part of a subhalo soliton is removed by tidal stripping, the subhalo soliton will re-equilibrate into a new configuration with a lower density and larger core radius. Such re-equilibration leads to a "run-away" effect that eventually disrupts the FDM subhalo soliton at a much faster rate than a CDM subhalo. Here, we compare all three of our simulated subhaloes with the results reported in \citet{Du2018}.

\begin{figure}
    \hspace{-1.4cm}
    \includegraphics[scale=0.35]{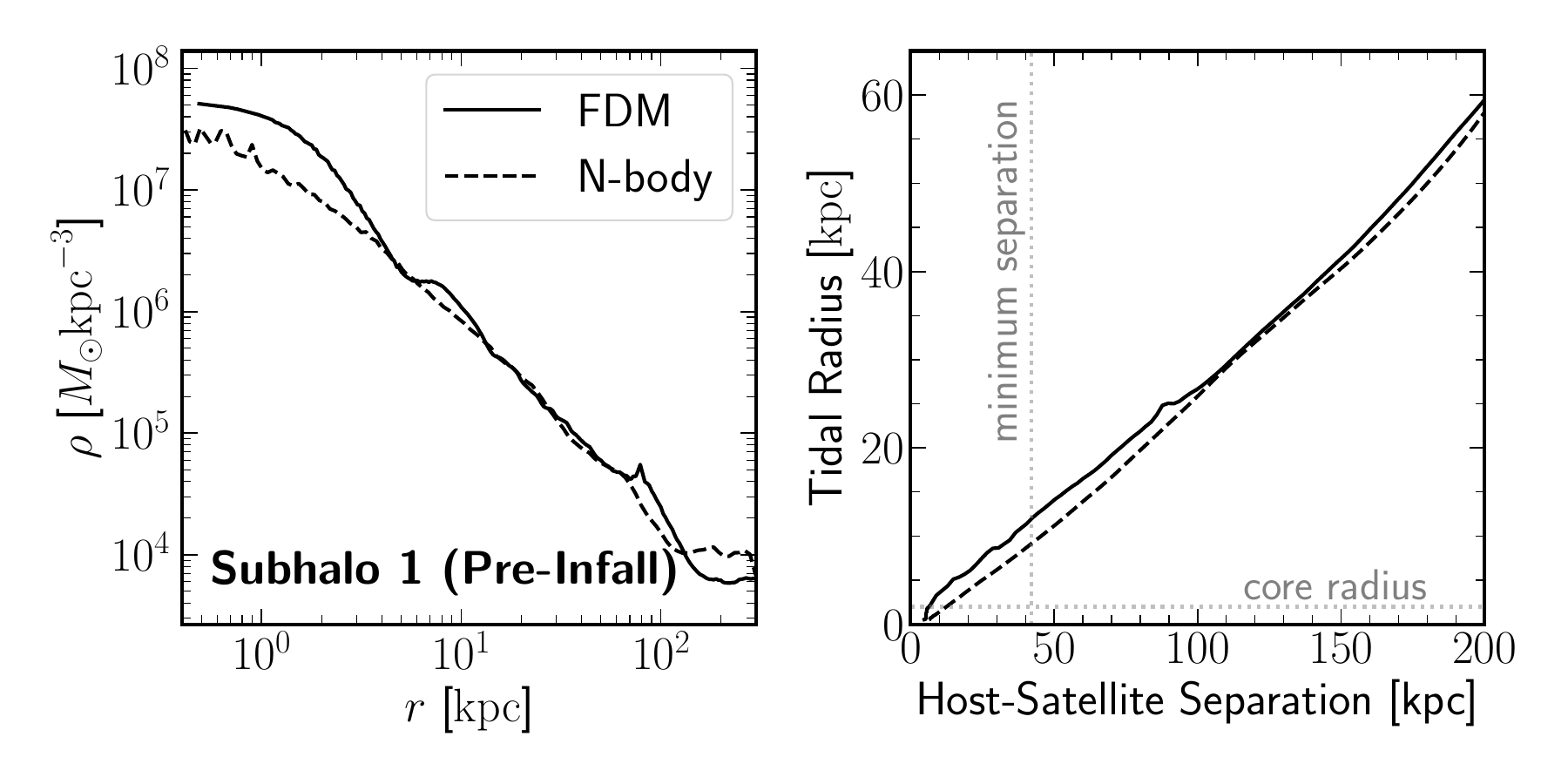}
    \caption{Comparison of subhalo 1 before infall between the FDM simulation (solid lines) and the N-body simulation (dashed lines). The FDM subhalo features a denser core than its N-body counterpart (left), resulting in a consistently larger tidal radius (right). Notably, the FDM tidal radius at the minimum host-satellite separation (vertical dotted line; see Fig. \ref{fig:subhalo_cdsgrr}) remains significantly larger than the core radius (horizontal dotted line), which explains why the FDM soliton is not tidally disrupted during pericenter passage.}
    \label{fig:tidal}
\end{figure}

\begin{figure*}
    \centering
    \hspace*{-1.0cm}
    \includegraphics[scale=1.5]{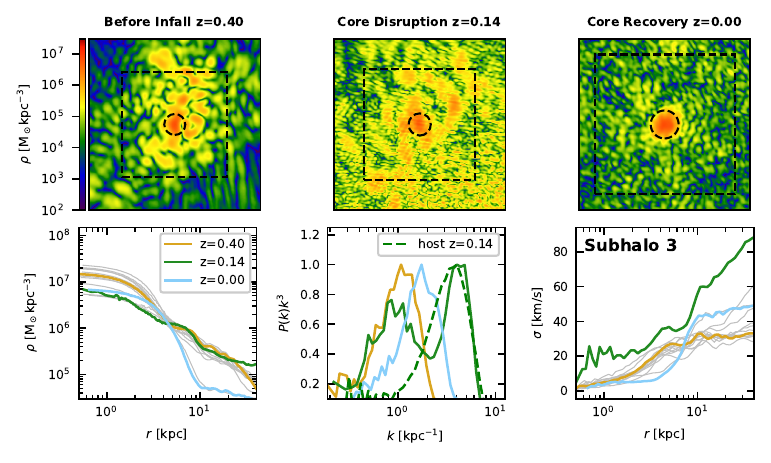}
    \caption{Properties of subhalo 3, which undergoes core disruption at redshift $z=0.14$ but recovers at $z=0$. At the moment of disruption, the subhalo is contaminated by interference fringes and the velocity dispersion profile is higher than any other point in its evolutionary history. The top row shows density slices at $z = 0.4, 0.14$, and $0$, with core radii of 2.44 kpc, 2.00 kpc, and 2.88 kpc, respectively. Since there is no apparent core structure at $z=0.14$, the core radius is given only for scale reference and calculating the power spectrum. See Fig. \ref{fig:subhalo} for details on each panel.}
    \label{fig:subhalo3}
\end{figure*}


The core of subhalo 3 in the FDM simulation was disrupted at $z=0.14$ but recovered close to $z\sim 0.0$. As shown in the top-middle panel of Fig. \ref{fig:subhalo3}, the core is contaminated by interference fringes at the time of disruption. This contamination indicates a perturbed soliton that is no longer in its ground state, likely due to energy injected from the host, as evidenced by the increased velocity dispersion of the soliton (see bottom-right panel of Fig. \ref{fig:subhalo3}). Despite this core disruption, subhalo 3 remains gravitationally bound after its pericenter passage. As it moves farther from the host center, the soliton re-emerges at $z=0.0$, where the influence of host granules is reduced (see top-right panel of Fig. \ref{fig:subhalo3}). This soliton re-emergence suggests that the thermalized core undergoes energy redistribution at the outskirts of the host halo, allowing the perturbed soliton to relax back to its ground state. Thermal arguments describing Bose-Einstein condensation of self-gravitating systems using kinetic theory have been applied in a non-cosmological framework \citep{Levkov2018, Chen&Marsh2021}. However, the evolution of subhaloes inside a host will change the outer envelope, potentially invalidating the assumption of thermal equilibrium used in those studies. Whether relaxation timescales are modified in such out-of-equilibrium systems remains an open question, and the soliton re-emergence warrants further investigation in future work.

\begin{figure*}
    \centering
    \includegraphics[scale=1.2]{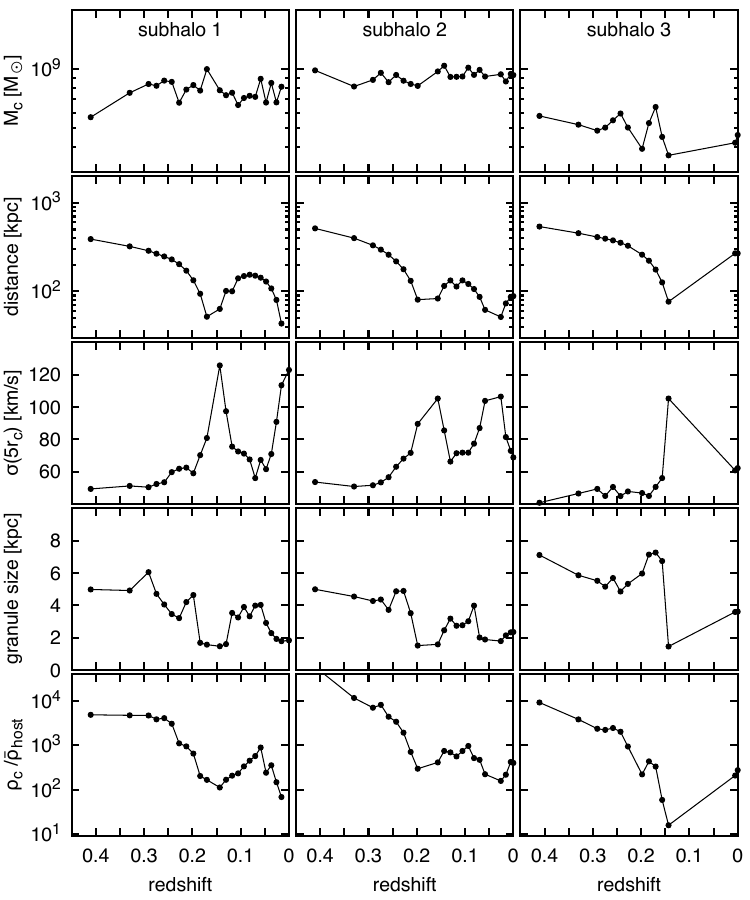}
    \caption{Evolution of core mass, distance to the center of the host, velocity dispersion at $5r_\mathrm{c}$, granule size, and core-host density ratio from $z=0.4$ to $0$ for all three subhaloes in the zoom-in simulation. Note that subhalo 3 (right column) has experienced tidal disruption around $z=0.14$ and re-emerged at $z\sim0$. }
    \label{fig:subhalo_cdsgrr}
\end{figure*}
\citet{Du2018} performed ideal simulations of FDM core disruption by including an external gravitational potential of the host halo, but a core filled with interference is not observed in their study, or other previous simulations as well, because of their difficulty in performing a self-consistent Milky Way-sized simulation.

\citet{Du2018} concluded that tidal disruption of an FDM soliton occurs when $\rho_\mathrm{c}/\bar{\rho}_\mathrm{host}<4.5$, where $\rho_\mathrm{c}$ is the central density of the soliton and $\bar{\rho}_\mathrm{host}$ is the average density of the host within the orbital radius. We examine their claim with our subhaloes, as shown in the top and bottom rows of Fig. \ref{fig:subhalo_cdsgrr}. It is clear that subhalo 1 and 2, which both have survived cores, are more massive than subhalo 3. Their $\rho_\mathrm{c}/\bar{\rho}_\mathrm{host}$ ratios remain above $\sim 100$ after infall, consistent with the condition for core disruption $\rho_\mathrm{c}/\bar{\rho}_\mathrm{host}<4.5$. In comparison, subhalo 3 is disrupted at $z=0.14$ when $\rho_\mathrm{c}/\bar{\rho}_\mathrm{host}=16$, which is slightly higher than the proposed threshold $4.5$. The underestimated threshold in \citet{Du2018} is possibly due to the lack of the background granules of the host in their simulations, which deposit additional kinetic energy and may enhance the instability of the soliton. We remind that when the remnant of the subhalo 3 distanced itself further from the host center, the core recovered at $z\sim0.0$ with a higher $\rho_\mathrm{c}/\bar{\rho}_\mathrm{host} \gtrsim 200$, which suggests the subhalo 3 is still gravitationally bound after the close encounter. 

The cores of subhalo 1 and 2 survive in the FDM simulation, but the corresponding subhalo 1 in the N-body simulation is disrupted. This result may appear contradictory to the expected enhanced tidal disruption in the FDM model at first glance. To understand this behavior, Fig. \ref{fig:tidal} presents the density profile of subhalo 1 immediately before entering the host. It is evident that the FDM subhalo features a denser core than its N-body counterpart due to the presence of a central soliton. Following \citet{Du:2016aik}, we estimate the tidal radius of subhalo 1 based on its density profile just before infall and the host halo's density profile at $z=0$. The results, as shown in the right panel of Fig. \ref{fig:tidal}, indicate that the tidal radius of subhalo 1 in the FDM simulation is consistently larger than in the N-body simulation. This larger tidal radius suggests that, despite the quantum tunneling effect, the deeper gravitational potential created by a compact, dense soliton helps protecting it from tidal disruption. In contrast, the shallower gravitational potential of subhalo 1 in the N-body simulation makes it more vulnerable to tidal stripping by the host, explaining its disruption in the N-body case but not in the FDM simulation. This finding further supports the presence of a denser core in FDM halos compared to CDM, consistent with the high-redshift findings \citep{Sandy2025}.

\subsection{Granules}
\label{sec:granules}
FDM model exhibits intricate interference patterns, which can only be properly resolved by the wave scheme but not by simulations only solving the Madelung equation. The interference pattern manifests itself as density granulation surrounding the solitonic core of each halo, and the granule size is of the order of de Broglie wavelength or the soliton size. In this work, we measure the granule size of the subhalo through its spectral peak in the density power spectrum, which is further verified by the velocity dispersion profile. Each power spectrum is computed from the 3D density grid of the subhalo within $|x-x_\mathrm{center}|,|y-y_\mathrm{center}|,|z-z_\mathrm{center}|<10r_\mathrm{c}$, where $(x_\mathrm{center}, y_\mathrm{center}, z_\mathrm{center})$ are the center of the subhalo soliton. The density grid is normalized by the density profile of the subhalo to put equal weight to different radii. Meanwhile, we exclude the central soliton by zeroing out the region within $r<2r_\mathrm{c}$. Such a method is similar to \citet{Chan2018}, but, instead of using a spherical shell, we measure the spectrum from a box-shaped density grid. In Appendix \ref{sec:Pk_test}, we verify the reliability of this method by testing it on an analytically generated density granulation with a pre-determined granule size. In this subsection, we demonstrate that the tidal evolution of a subhalo can be traced by its granule size.

\subsubsection{Power Spectra}
Fig. \ref{fig:subhalo} highlights three particular redshifts to demonstrate the tidal evolution of subhaloes by using the granule size changes of subhalo 2. Before infall, the subhalo behaves as an isolated halo outside of the host, so we expect the granules to have sizes comparable to the subhalo soliton. The power spectrum at $z=0.4$ (brown line) peaks at $k_\mathrm{max}=1.26~\mathrm{kpc}^{-1}$, corresponding to a granule size of $2\pi/k_\mathrm{max}=5~\mathrm{kpc}$. Since the subhalo before infall has a core radius of $r_\mathrm{c}=1.34~\mathrm{kpc}$, our measured granule size is approximately $2$ times the core diameter, which is consistent with the fact that $2.2r_\mathrm{c}$ encloses $90$ percent of the total soliton mass \citep{Chan2018}. 

After infall, the spectral peaks at $z=0.2$ (green) and $z=0$ (blue) are $k_\mathrm{max}=1.76~\mathrm{kpc}^{-1}$ and $2.64~\mathrm{kpc}^{-1}$ respectively, which both have smaller granule sizes than that before infall. The spectral peak of the host (dashed blue line in the bottom-middle panel of Fig. \ref{fig:subhalo}) is at $k_\mathrm{max}=3.20~\mathrm{kpc}$\footnote{The spectrum of the host is measured using the same method except that the 3D density grid centers at a region without any subhalo while maintaining the same distance from the center of the host.}. 
The changes in the granule size of subhalo 2 can be explained by its interaction with the host. As the subhalo approaches the center of the host, the granules of the subhalo are in superposition with the granules of the host. The latter gradually dominate over the former until the subhalo is fully stripped. Therefore, when measuring the granule size of the subhalo during its earlier stages of infall, we capture the granules of the subhalo itself, or in superposition with the host. Once the subhalo is significantly stripped, despite the measurement being taken to be close to the core of the subhalo, it reflects the granule size of the host. As expected, the subhalo can only reach a certain minimum granule size, close to $2~\mathrm{kpc}$, corresponding to the granule size of the host halo.  

For $m>2\times 10^{-23}~\mathrm{eV}$, the dominance of host granules over stripped subhaloes is also expected in theory. The key question is the extent to which the host granules dominate over the subhalo volume. As the soliton radius scales as $m^{-1}$ for a fixed subhalo mass, the subhalo granules with heavier particle masses will extend further, occupying regions that were originally part of the soliton cores in our simulation. Moreover, since the soliton mass also scales as $m^{-1}$ for a fixed subhalo mass, the subhalo gravity around the soliton radius increases linearly with $m$. As a result, the ambient medium surrounding the soliton becomes less vulnerable to tidal stripping for heavier particle masses. Quantitatively addressing the dominance, such as comparing the soliton radius to the extent of the region dominated by subhalo granules, requires further simulations of various particle masses.

Such a result is important for constraining FDM particle mass using dynamical heating effects caused by soliton random motion or fluctuating granules. For instance, \citet{Marsh2019} constrained FDM particle mass through the survival of the star cluster within Eridanus II, considering the effects of soliton oscillations and fluctuating granules. \citet{Schive2020} later revealed that soliton random walk can be greatly reduced by stripping away the granules outside the soliton. However, our results directly demonstrate that a tidally stripped subhalo retains a soliton surrounded by the granules of the host halo, instead of a naked soliton without granules. Since soliton oscillations and random walk can be understood as the superposition of the ground state (the soliton) and excited states (the granules), we expect the soliton of a tidally stripped subhalo to exhibit oscillations and random walk due to the superposition with the excited states of the host halo \citep{Li2021}. It will be important to revisit the quantitative analysis of the frequency and amplitude of soliton random walk and oscillations in this more realistic scenario when more simulated subhalo samples become available. In addition, \citet{Dalal2022} have made constraints on FDM particle mass based on the dynamical heating effect from the granules of the subhalo on ultrafaint dwarf galaxies. Since our self-consistent subhalo has confirmed the decreasing granule size within the host, it will be necessary to revisit the heating rates accounting for the presence of the host halo in future analyses.  


Intriguingly, the power spectrum of subhalo 3 at $z=0.14$ shows two peaks (see Fig. \ref{fig:subhalo3}), with one occurring at lower $k$ corresponding to the granule size of the subhalo, and the other at higher $k$ corresponding to the granule size of the host halo. We confirm the above by showing the power spectrum of the host at $z=0.14$ (green dashed line), which peaks at a similar $k$ value to the higher-k peak of the subhalo (green line). It reveals that the granules of a subhalo are superposed with that of the host at the moment of core disruption. Since the granules of subhalo 3 were larger, due to its smaller halo mass, we can observe two distinct granule sizes reflected by the two distinct peaks in the power spectrum.

\subsubsection{Velocity Dispersion}

The granule size inversely scales with the de Broglie wavelength, $\lambda=h/(m\sigma)$, where $\sigma$ is the local velocity dispersion. Thus, we can verify the measured granule size, as discussed in the previous subsection, by comparing it with the velocity dispersion of the subhalo.

The lower-right panel of Fig. \ref{fig:subhalo} shows the time-evolving velocity dispersion profile of subhalo 2 during its infall.
Similar to the velocity dispersion profiles of the host halo, the subhalo velocity dispersion is small within the soliton for all redshifts, consistent with the fact that the soliton is supported by quantum pressure rather than turbulent motion. Beyond $\sim 2r_\mathrm{c}$, the velocity dispersion remains constant at $z=0.4$ but shows a radial increase at $z=0.2$, and $0.0$.

To fully reveal the relation between granule size, velocity dispersion and distance to the center of the host, we show the evolution of these quantities as a function of redshift in Fig. \ref{fig:subhalo_cdsgrr}. As subhalo 2 gets closer to the center of the host, the velocity dispersion increases while the granule size decreases throughout its entire evolution after the infall, which is consistent with the inverse scaling relation of the de Broglie wavelength. Again, the results can be interpreted as subhalo 2 being tidally stripped by the host halo; the superposition of the subhalo and host halo wave functions results in the shrinking granule size. The velocity dispersion of subhalo 1 follows a similar evolution.

For subhalo 3, the sudden decrease in the granule size and increase in the velocity dispersion in Fig. \ref{fig:subhalo_cdsgrr} signify the moment of core disruption at $z=0.14$. The bottom-right panel in Fig. \ref{fig:subhalo3} shows that subhalo 3 has a high velocity dispersion even within the core at $z=0.14$, contradictory to the low velocity dispersion expected inside an intact soliton. 

This fact confirms that the core of subhalo 3 is primarily disrupted by tidal stripping from the host, with granule heating further enhancing the instability of its quantum pressure-supported core. Notably, the velocity dispersion decreases again as the subhalo 3 moves farther from the center of the host, indicating the recovery of the soliton.

\section{conclusion}
\label{section:conclusion}
This work presents a cosmological FDM simulation of a Milky Way-sized halo in a $14.85~\mathrm{cMpc}$ box using the improved \texttt{GAMER-2} code. The recently implemented novel hybrid scheme solves the Hamilton-Jacobi-Madelung equation on large scales while adopting a highly accurate local spectral solver for the Schr\"odinger equation to resolve non-linear wave dynamics on small scales. By combining the AMR algorithm, hybrid scheme, and zoom-in technique, it now becomes feasible to resolve FDM substructures within a Milky Way-sized halo in a self-consistent way. Although our simulation is performed with a light particle mass $m=2\times10^{-23}~\mathrm{eV}$, it can fully reveal the FDM subhaloes going through each evolutionary stage: from the formation of an isolated halo to becoming tidally stripped and disrupted. Three subhaloes survive at z = 0 in our simulated host halo, and our main findings can be summarized as follows:\\ 

$\bullet$ The density profiles of the simulated FDM Milky Way-sized host halo match with its N-body simulated counterpart at large radii, confirming numerical convergence. The velocity dispersion profiles of the host vanish at the center of the core, which is consistent with the minimal kinetic energy content in a quantum pressure-supported solitonic core. However, such a velocity profile is in contrast to the FDM halo simulated by the Smooth Particle Hydrodynamic approach in \citet{Nori2023}. The difference in the solitonic core between different numerical methods remains to be further studied. 

$\bullet$ We observe the core mass of tidally stripped subhaloes remain the same size after infall, suggesting that the core mass-halo mass relation from previous simulations \citep{Schive2014b, Liao2024} is inapplicable to those of tidally stripped subhaloes. In other words, it is recommended to adopt the halo mass at the time of infall to model the core of a tidally stripped FDM subhalo.

$\bullet$ The granulation of the host halo plays an important role in the tidal evolution of FDM subhaloes. Our tidally stripped subhaloes do not have a naked soliton without granules, but rather a soliton surrounded by the granules of the host halo.  The granule size of the subhaloes correlates with the distance to the center of the host, demonstrating how the host granules gradually dominate the subhalo granules until they are fully stripped. This finding has implications for existing observational constraints on the FDM particle mass based on soliton oscillations and random motion, such as those made by \citet{Marsh2019}, \citet{Schive2020} and \citet{Dalal2022}, which did not account for the realistic scenario of a subhalo surrounded by host granules. The extent to which the amplitude of these dynamical heating effects is affected by the presence of surrounding host granules will be investigated in future studies with larger samples of simulated subhaloes.

$\bullet$ We observe the subhalo core is contaminated by interference during core disruption, which reflects granule heating of the subhalo core by the granules of the host. Such a heating effect could enhance core instability and suggests a higher core disruption threshold than the criterion $\rho_\mathrm{c}/\bar{\rho}_\mathrm{host}=4.5$ proposed by \citet{Du2018}.

$\bullet$ FDM subhaloes have a denser core than their N-body counterparts immediately before infall, consistent with the findings in \citet{Sandy2025}. This denser pre-infall core creates a deeper gravitational potential, making FDM subhaloes more resistant to tidal disruption.

The particle mass $m=2\times10^{-23}~\mathrm{eV}$ adopted in this work is disfavored by various observations \citep[e.g.][]{Nadler2021,Rogers:2020ltq,Dalal2022}. However, the conclusions on the halo-subhalo interactions in a cosmological context, particularly the constant core mass during infall and the presence of surrounding host granules in a tidally stripped subhalo, are expected to remain valid for heavier particle masses. Moreover, this work demonstrates the potential of using the newly developed hybrid scheme in \texttt{GAMER-2} for future work to perform more simulations of Milky Way-sized haloes with different FDM particle masses, obtain larger samples of subhaloes, and conduct more statistically robust analyses of stripped FDM subhaloes. We will explore these directions in future work.

While this study does not involve hydrodynamic simulations, we expect baryonic feedback to influence our simulated subhalo mass, which is $\sim 10^{11}~\mathrm{M}_\odot$ before infall and lies within the range affected by baryonic feedback \citep{Tollet2016}. Detailed studies on the impact of baryonic feedback in the FDM model remain limited \citep{velmatt2020,mocz2019}. However, a recent study by \citet{robles2024} demonstrated the effect of supernova feedback on FDM solitons. Our subhaloes are surrounded by density granules, leading to intrinsic oscillations of the subhalo cores. According to \citet{robles2024}, supernova feedback may inject additional kinetic energy into solitons, thereby amplifying these stochastic soliton oscillations.

There exist other substructures in our simulations, such as streams, that are waiting to be found. However, our current primitive subhalo finder cannot identify stream-like structures. We are also unable to identify the edge of subhaloes, which defines the subhalo mass. Given that self-consistent simulations of substructures are now more feasible for the FDM model, it is worthwhile to start investing effort in developing a more sophisticated substructure finder tailored for the FDM model. 

\begin{figure*}
    \centering
    \includegraphics[scale=0.6]{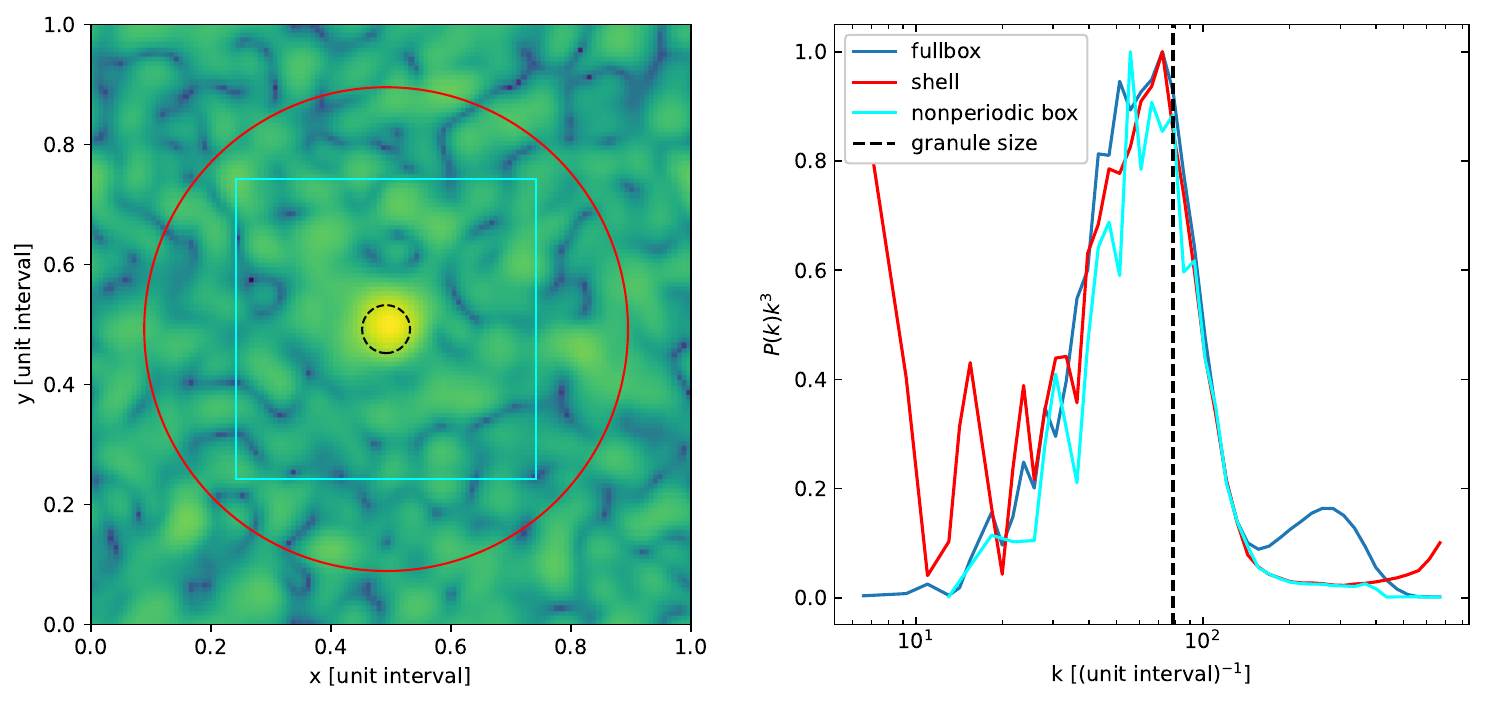}
    \caption{A $128 \times 128$ slice of analytically generated density granulation within a unit square (left) and the measured power spectra (right) using different kinds of data sets.}
    \label{fig:Pktest}
\end{figure*}

\section*{Acknowledgements}
We thank Xialong Du for providing the fitting parameters for the subhalo mass function. We would also like to thank Tzihong Chiueh, Elisa G. M. Ferreira, and Andrew Eberhardt for extensive discussions related to this work. This work was supported by the National Science and Technology Council (NSTC), the Ministry of Education (Higher Education Sprout Project NTU-113L104022-1), and the National Center for Theoretical Sciences (NCTS) of Taiwan under Grant No. 111-2124-M-002-013, 112-2124-M-002-003, and 113-2124-M-002-003. We thank to National Center for High-performance Computing (NCHC) for providing computational and storage resources. HYS acknowledges support from NSTC under Grant No. NSTC 111-2628-M-002-005-MY4 and the NTU Academic Research-Career Development Project under Grant No. NTU-CDP-113L7729.
The following software and libraries were used for data analysis and visualization: \texttt{NumPy} \citep{NumPy}, \texttt{Matplotlib} \citep{matplotlib}, \texttt{scikit-learn} \citep{scikit-learn}, and \texttt{yt} \citep{ytproject}.

\section*{DATA AVAILABILITY}
The data underlying this article will be shared on reasonable request to the corresponding author. 


\input{references2.bbl}




\appendix

\section{Power Spectrum Test}
\label{sec:Pk_test}
We follow a similar procedure from \citet{Chan2018} to measure the size of granules using the peak of the power spectrum. The left panel of Fig. \ref{fig:Pktest} shows a slice of a 3D density granulation within a cube of side length $1$, generated by 
a velocity dispersion that assumes a granule size of $0.08$ in the unit interval. The corresponding wavenumber is $78.5$ per unit interval. We first normalize the 3D density distribution by dividing it with the spherically averaged density profile $\rho(r)$ centered at the soliton. With the known granule size, we test it against power spectra measured in three different kinds of data: unfiltered data (blue), a spherical shell of density (red), and an outer box-shaped shell while excluding the central soliton (cyan). For the case of a spherical shell, we zero out the data outside of the shell, whereas for the case of a box-shaped shell, we trim the data into a smaller non-periodic box. 

The resulting power spectra are shown in the right panel of Fig. \ref{fig:Pktest}. We see that all three kinds of data sets produce spectral peaks in agreement with our pre-determined granule size. Therefore, this test demonstrates that all three methods are equally suitable for measuring the granule size of FDM haloes, with some subtleties. The unfiltered data set produces a high-k bump in the power spectrum which comes from the size of the soliton. Also, the spherical shell data fluctuates in the low-k scale of the power spectrum. As such, we conclude that the trimmed box-shaped shell data set gives the cleanest power spectra, where the high-k bump and low-k fluctuation are absent. 
\section{Abundance of Subhaloes}
\label{sec:abundance}

In total, three subhaloes survive until $z=0$ in our simulation, which all can be seen in the projected density in Fig. \ref{fig:MWz0}. The abundance of subhaloes in our simulation can be tested against the semi-analytical FDM subhalo mass function developed by \citet{Duthesis}. The model constructs the subhalo mass function and merger trees using the Press-Schechter formalism while accounting for the small-scale suppression and the tidal effects of the FDM model. The fitting formula is as follows:
\begin{equation}
\label{Eq:Du2017}
\begin{aligned}
\left.\frac{dN}{d\mathrm{ln}M}\right|_\mathrm{FDM} &= \beta\mathrm{exp}\left[-\left(\mathrm{ln}\frac{M}{M_1\times10^8M_\odot}\right)^2/\gamma\right] \\
& \hspace{-0.5 cm} + \left[1+\left(\frac{M}{M_2\times10^8M_\odot}\right)^{-\alpha_1}\right]^{10/\alpha_2}\left.\frac{dN}{d\mathrm{ln}M}\right|_\mathrm{CDM}.
\end{aligned}
\end{equation}
We adopt the following parameters provided by the author of \citet{Duthesis}: $M_1=42.667$, $M_2=24.458$, $\alpha_1=0.9121$, $\alpha_2=0.71186$, $\gamma=1.13077$ and $\beta=5.1021\times10^{-4}$ for $m=2\times10^{-23}~\mathrm{eV}$.

We remind that we cannot use the virial criterion to define the subhalo boundaries and the corresponding subhalo mass $M_\mathrm{sub}$ in our simulation because we currently cannot isolate the subhalo from the background of the host halo.  However, the semi-analytical model stops the tidal stripping effect once their subhalo bound mass reaches $4M_\mathrm{c}$, which can be interpreted as the minimum subhalo mass in their model. Accordingly, we define subhalo mass as $4M_\mathrm{c}$ as well for our simulated subhaloes to allow a fair comparison. Note that the total soliton mass without an outer halo envelope is also $\sim 4M_\mathrm{c}$ \citep{Schive2014a}, so a partially tidally stripped subhalo will have a larger subhalo mass than $4M_\mathrm{c}$. As a result, the definition of subhalo mass here can underestimate its true value.

\begin{figure}
    \centering
    \hspace*{-0.8cm}    \includegraphics[scale=0.48]{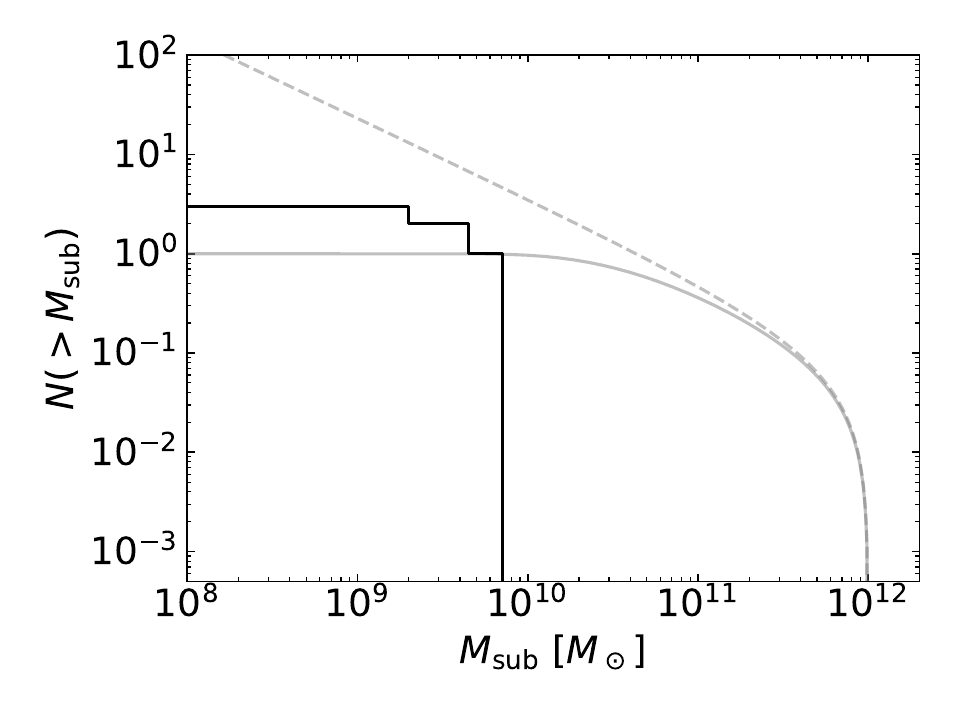}
    \caption{Cumulative subhalo mass function at $z=0$ in a Milky Way-sized host for $m=2\times10^{-23}~\mathrm{eV}$. The subhalo mass function of this work truncates at a smaller subhalo mass while having a higher total number of subhaloes than the semi-analytical model of \citet{Duthesis}. 
    }
    \label{fig:cshmf}
\end{figure}

Fig. \ref{fig:cshmf} shows the cumulative subhalo mass function by integrating Eq. (\ref{Eq:Du2017}), predicting a maximum of one subhalo in a Milky Way-sized host with $m=2\times10^{-23}~\mathrm{eV}$. In comparison, our simulation shows three subhaloes in a Milky Way-sized host halo.  In addition, the semi-analytical model only starts to drop around $M_\mathrm{sub}\sim 4\times10^{10}~M_\odot$, corresponding to the peak of the subhalo mass function, whereas our simulated host does not contain any subhalo with $M_\mathrm{sub}>7\times10^9~M_\odot$. Therefore, our simulated host halo contains more subhaloes, but they are less massive compared to the semi-analytical model. We stress that a more rigorous check of the assumptions in Eq. (\ref{Eq:Du2017}) and additional simulations of Milky Way-sized FDM haloes are needed to improve the statistical robustness in addressing this mild tension.

\bsp	
\label{lastpage}
\end{document}